\documentclass[runningheads]{llncs}
\usepackage[T1]{fontenc}
\usepackage{graphicx}
\begin{document}
\title{Predictive Process Monitoring: a comparison survey between different type of event logs}
%
%
\author{Simona Fioretto\inst{1}\orcidID{0009-0006-8700-8188} \and
Elio Masciari\inst{1}\orcidID{0000-0002-1778-5321}}
\authorrunning{S. Fioretto et al.}
%
\institute{University of Napoli Federico II, Napoli, Italy
\email{\{simona.fioretto,elio.masciari\}@unina.it}}

\maketitle              

\begin{abstract}
The application of Predictive Process Monitoring (PPM) techniques is becoming increasingly widespread due to their capacity to provide organizations with accurate predictions regarding the future behavior of business processes, thereby facilitating more informed decision-making. A plethora of solutions have been proposed in the literature employing these techniques, yet they differ from one another due to a number of factors. However, in light of the growing recognition of the value of object-centric event logs, including in the context of PPM, this survey focuses on the differences among PPM techniques employed with different event logs, namely traditional event logs and object-centric event logs. In addition, the reviewed methods are classified according to the prediction task they address and the specific methodologies they employ.

\keywords{Process Mining  \and Predictive Process Monitoring \and Object-Centric Event Log.}
\end{abstract}
\section{Introduction}
In the context of the ongoing digital transformation, the growing volume of operational data stored within information systems is emerging as a significant business opportunity, offering potential for the development of advanced applications such as information retrieval. 

Event records of processes execution widely known as Event Logs \cite{augusto2018automated} can reveal important insights about business processes, thus enabling the development and application of Process Mining (PM) research area \cite{van2016data}.  Moreover, the rising adoption of Statistical Models, Machine Learning techniques, and AI in business solutions \cite{ali2020artificial} contributes to the advancement of PM area.

The process of managing business processes is defined by the field of Business Process Management (BPM) \cite{dumas2018fundamentals}. The management of a business process is a cyclical process, beginning with the identification of the process itself, followed by the discovery and analysis of the process, then the redesign and implementation of the process, and finally the monitoring of the process. 
\emph{Process monitoring} enables real-time extraction of information on the actual execution of the process, facilitating the acquisition of knowledge on process measures and performance, and identifying potential problems. 

The monitoring phase can be conducted in either an offline or online mode, thus enabling a transition from the application of a-posteriori techniques based on historical data, to the utilization of runtime techniques while the process is ongoing.

This paper focuses on Predictive Process Monitoring (PPM) approaches \cite{dumas2013process}, which support online process monitoring within the context of Process Mining. PPM offers insight into the future characteristics of ongoing cases by training a prediction model on the event log data of completed cases.

In this survey, we focus on the identification of PPM methodologies based on the specific input event logs that are being processed. In this context, we differentiate between two primary categories: the first encompasses traditional event logs, and the second, object-centric event logs, which, to the best of our knowledge, has not been previously considered. 
This study elucidates the methodologies deployed to integrate object-centric event log data into classical event log PPM methodologies. Through a comprehensive analysis, it demonstrates the significance of utilising this information to enhance prediction accuracy. Furthermore, it provides insights into the potential of adapting existing methodologies for novel applications.

The paper is structured as follow: first, an overview of the meaning of predictive process monitoring will be provided, and the difference between an object-centric event log and a classical event log will be highlighted in Section \ref{sec:preliminaries}. Then it will be provided a classification of the existing surveys in Section \ref{sec:related_work}. In Section \ref{sec:search_methodology} the methodology employed for the retrieval of the studies is analysed, as is the classification system used to categorise them in Section \ref{sec:methods-classification}. 

\section{Overview} \label{sec:preliminaries}
In this section we provide an overview of the concepts of Event Log and Predictive Process Monitoring. 

\subsection{Predictive Process Monitoring}
Predictive Process Monitoring (PPM) is a specific research area of Process Mining (PM) that employs historical complete process execution traces to make predictions about the future evolution of features of an uncompleted execution trace of an ongoing process \cite{maggi2014predictive,DiFrancescomarino2022}. By leveraging PPM it is possible to predict the performance indicators of a running process, such as the outcome of a process execution, its completion time, or the sequence of its future activities to cite a few. 
Event Log traces, i.e. completed process executions are used to train a Predictive Model, which is in general a Machine Learning model trained on historical data X to assign a label on new data Y in the context of prediction. The main objective of the model is to generalize the pattern identified in the training data, providing a mapping X → Y. \cite{verenich2019survey}.More in detail, given the previous concepts, it is then possible to define the general PPM framework. This framework starts from event logs data containing information on the history of the process (i.e., historical complete process executions) which are given as input for the specific prediction algorithm to be implemented. The algorithm is trained to discover patterns in data and then generalize them. Finally, during the execution of the process, the previously trained model takes as input the event stream or prefix logs, which are live data representing the ongoing trace of a runtime process, to perform the prediction task, thus supporting decisions about the ongoing and future execution.

\subsection{Event Log}
Process Mining approaches rely on the availability of Event Logs in Information Systems, data in which the execution of processes is recorded \cite{van2012process,van2019object}. 

The Event Log utilized in the context of Process Mining has a specific structure.  In order to be utilized, it needs to have three mandatory fields about the process execution. These are respectively \begin{itemize}
    \item the specific process instance, called \emph{case};
    \item the actual action being recorded, called \emph{activity};
    \item the \emph{timestamp} of the activity being recorded;
    \item optional \emph{attributes} that could be recorded during the execution of the process (i.e., used resources, involved people, incurred costs, etc.)
\end{itemize}

The aforementioned specification includes the existence of a unique case notion, with each event linked to exactly one case identifier \cite{van2019object}. However, in real-world organizational scenarios, this assumption may be insufficient to properly model the data \cite{van2019object,van2020discovering,van2021object}. To address this \emph{Object-Centric Event Logs} are introduced, to ease the previous constraints.

The importance of enhancing data quality and reliability in process mining applications might pursue integrity checking practices ~\cite{CM:LOPSTR2003,CM:FoIKS2004,M:FQAS2004,M:PHD2005,DM:LPAR2006,DM:FlexDBIST2007,DM:FlexDBIST2006,MC:DEXA2005},
top-$k$ selections ~\cite{MT:TKDE2011,CM:CIKM2018,CM:TODS2020,MT:PVLDB2010,CCFMT:TODS2013} to retain the most promising outcomes, or even leverage collective input from multiple stakeholders to accurately map out processes as in crowd sourcing ~\cite{DBLP:conf/mmsys/LoniMGGMAMMVL13,DBLP:conf/socialcom/GalliFMTN12,DBLP:conf/www/BozzonCCFMT12}.
We also observe that accessing process mining logs poses challenges similar to those that were addressed in the field of query processing under access limitations~\cite{CM:ER2008,CCM:JUCS2009}.

Object-Centric Event Logs are used to model and preserve information from real-world scenarios, where data often take the form of graph structures characterized by the interaction and synchronization of multiple objects. This means that a single business process cannot be tied to just one object, such as a product or service. Instead, various objects (e.g., client orders, shipment orders, product IDs) are interconnected through shared activities. These interconnections give rise to complex graph structures that reflect the diverse nature of business processes. Object-Centric Event Logs assume the existence of multiple case notions, or object types, and allow an event to reference multiple objects across different types \cite{van2019object}. In contrast to Classical Event Logs, where each object has a unique case ID and represents a distinct process, Object-Centric Event Logs account for shared activities and structural similarities between processes, linking individual processes to those of other objects. Thus, the behavior of OCEL processes is defined by the synchronization of the life cycles of these objects.

\section{Related Work} \label{sec:related_work}
This section presents an overview of the surveys proposed and the benchmarking carried out in the PPM area. In \cite{van2016data} Van der Aalst present the most comprehensive manual on process mining, providing deep details of Process Mining research are.  In their survey on process discovery, the authors of this study \cite{augusto2018automated} identify this as one of the most significant challenges in process mining. In \cite{carmona2018conformance}, the authors conducted a survey on the various approaches employed in the context of conformance checking, thereby further contributing to the field's body of knowledge.

However, in this section we focus on comprehensive studies that can guide the reader in an complete introduction of the specific PPM field.

Marquez et al. \cite{marquez2017predictive} reviewed 39 PPM studies, categorizing approaches into process-aware and non-process-aware, by output type (numerical/categorical) and by prediction task (regression/classification). Input data were classified according to the perspective as control-flow or data-flow. Key findings include limitations of scarce datasets and reliance on feature vectors or full models, reducing generalizability, lack of benchmarks, and the need for interpretable models. 

Di Francesco Marino et al. \cite{di2018predictive} analyzed 51 studies, categorizing predictions by goal (time, outcomes, risks, etc.), input data (event logs with contextual enrichment), and methods (regression, neural networks, clustering). Then they examined the application domains, model performance (precision, recall, F-measure), and the importance of aligning techniques with business needs to enhance efficiency and guide research.  

Teinemaa et al. \cite{teinemaa2019outcome} reviewed and benchmarked outcome-oriented methods, focusing on offline (training) and online (real-time prediction) phases. Key offline activities included prefix selection, trace bucketing, sequence encoding, and classification algorithms. Their benchmark on 7 primary studies evaluated prediction accuracy, earliness, and computation time.  

Verenich et al. \cite{verenich2019survey} examined 24 studies on remaining time prediction, distinguishing discriminative (non-process-aware) and generative (process-aware) models. Prefix bucketing and encoding were evaluated for improving predictions. A benchmark using 16 event datasets evaluated accuracy (via MAE) and earliness, with training/testing split temporally and hyperparameters optimized through grid search.  

Tama et al. \cite{tama2019empirical} performed a benchmark of ML techniques for next-event prediction, on 6 public datasets, 20 classifiers, and various validation methods (cross-validation, hold-out). Sampling included case-based and event-based prefixes. The Credal Decision Tree (C-DT) consistently achieved the best accuracy, performing well across datasets and showing resilience to variability, particularly in high-variability logs.  

Harane et al. \cite{Harane2020} reviewed deep learning (DL) approaches in PPM, categorizing studies by network type (LSTM, Stacked Autoencoders), encoding methods (Word Embedding, One-Hot, N-grams), and prediction type (numerical, categorical, next-event). The survey highlighted DL’s effectiveness but emphasized weaknesses like limited datasets and a need for improved interpretability.  

Rama et al. \cite{rama2021deep} did a review and a benchmark on DL techniques in PPM, focusing on input data, sequence encoding, and neural network features. 10 approaches were evaluated on 12 datasets using Bayesian analysis. Insights highlighted model applicability and the trade-off between simplicity and performance.  

Stierle et al. \cite{Stierle2021BringingLI} reviewed 19 explainable PPM approaches, focusing on XAI concepts (intrinsic models, post-hoc explanations). Recent trends show a shift towards DL methods, though intrinsic interpretability dominated pre-2020. Evaluation of explainability methods remains a challenge.  

Neu et al. \cite{neu2022systematic} reviewed 32 studies on DL methods, analyzing network architecture (FFNN, CNN, RNN), preprocessing (one-hot, embeddings), and prediction targets (horizon, type). Challenges include input size constraints and case-specific temporal data encoding, with improvements needed in preprocessing and input data.

\section{Research Method} \label{sec:search_methodology}
This section presents the details of the phases involved in the execution of our Systematic Literature Review (SLR), in accordance with the steps proposed by \cite{kitchenham2004procedures}.

\subsection{Review Questions}
The initial stage of the SLR process entails the identification of the objective and the formulation of the research question.  
The rationale behind this SLR is the limited existing research on the impact of incorporating information from object-centric event logs in order to enhance predictions in Predictive Process Monitoring. Indeed, given the increasing prominence of object-centric event logs, there is a need for a more structured approach to research. This should address how to integrate such information, and how it affects and is affected by the broader variables in the research context, including the predictive task and the method employed. These considerations led to the formulation of the following research questions: \begin{itemize}
    \item [] $RQ_1$: What is the nature of the input event logs? 
    \item [] $RQ_2$:  What constitute the predictive tasks?
    \item [] $RQ_3$:  Which methods are employed in PPM?
\end{itemize}

The aforementioned research questions guided the analysis conducted in the present study.

\subsection{Study Selection}
The study selection process comprised the formulation of queries aligned with the research questions, designed to retrieve literature on the integration of object-centric event logs in predictive process monitoring (PPM). The queries employed were 'predictive process monitoring", "process mining", and "business process". These were executed across Scopus, IEEE Xplore, and Google Scholar with the objective of capturing studies on process monitoring, mining, and business process management. The analysis of Scopus data revealed an increase in the number of publications and citations in the period following 2019, indicating a focus on recent studies from 2019 to 2023. In order to be included in the study, approaches to PPM had to make use of machine learning, AI, statistical methods, or diverse event logs. In order to be excluded from the study, studies had to be unrelated to PPM, predictions had to be outside the scope of process mining, or the focus of the study had to be solely on explainability. 

A total of 135 results were returned by the initial search on Scopus. Following the application of filters to ensure the relevance and citation thresholds were met, this number was reduced to 61. A search on Google Scholar returned 521 results, which were then refined to 47 after a review of the top 10 pages. A search of the IEEE Xplore database identified 12 studies that were deemed to be relevant to the topic under investigation. Following the removal of duplicates and irrelevant content, a total of 25 articles were selected for analysis.

\section{Research Methodology} \label{sec:methods-classification}
According to the results and the research questions, the obtained studies can be categorized by the following dimensions: \begin{itemize}
    \item \emph{Event Log Type}: the type of event log considered for the application of the PPM is used to distinguish among the approaches; 
    \item \emph{Prediction Task}: we classify the considered papers based on the type of prediction;
    \item \emph{Method}: this dimension takes into account the predictive model being leveraged;
\end{itemize}

The selected papers will be discussed w.r.t. the above mentioned dimensions in what follows.

\paragraph{Event Log Type}
Most current PPM approaches assume a single case ID in classical event logs. However, real-life business processes often interconnect, challenging this assumption. This limitation led to the introduction of object-centric event logs (Object-Centric Event Log). Unlike standalone processes, this paradigm highlights interactions among objects (case notions), with concurrent process instances influencing prediction algorithms. Object-Centric Event Logs provide richer process execution data than classical event logs, impacting prediction tasks. Consequently, various PPM approaches emerged to explore the potential of Object-Centric Event Logs for improving prediction accuracy. Comparing approaches for classical and object-centric event logs helps clarify how researchers tackled these logs, the results achieved, and their relation to classical PPM methods.

\paragraph{Prediction Task}
According to \cite{DiFrancescomarino2022}, prediction tasks can be categorized into three types: outcome-based prediction, numeric value prediction, and next event prediction. These tasks significantly influence the choice of prediction methods. By incorporating this aspect into the classification, it becomes possible to determine whether certain prediction tasks are more common, if they rely on specific methods, and how they address differences between types of event logs.

\paragraph{Method} 
PPM methodologies fall into two categories: model-based approaches, which explicitly represent processes, and non-process-aware approaches, which rely on implicit representations. Non-process-aware methods, such as Machine Learning (ML) and Deep Learning (DL), are the most common. ML uses statistical models for predictions, while DL employs deep neural networks for more complex predictions. The choice of method depends on factors like event log characteristics and predictive tasks. It also balances features like interpretability, explainability, training time, and accuracy. DL methods often prioritize accuracy but lack explainability and require significant training time, making them less suitable for on-site use. Conversely, ML methods offer explainable but less accurate predictions. Each approach balances accuracy, explainability, and training time differently. Including the \emph{Method} feature in the classification allows identifying the most common methods and the aspects they prioritize: accuracy, explainability, or efficiency.

\section{Predictive Process Monitoring Approaches}
In the following sections the examined proposals are shown in detail by dividing them according to the Event Log Type. 

\subsection{Object-Centric Event Log}
The PPM approaches for Object-Centric Event Logs are recent and continuously evolving, resulting in limited but varied proposals. This section reviews PPM approaches for OCEL, detailing methods, prediction tasks, and contributions.

Gherissi et al. \cite{gherissi2022object} proposed a PPM methodology using LSTM networks to predict the next activity, next event time, and remaining time in Object-Centric Event Logs. Logs are preprocessed via flattening based on a chosen object type, generating enriched logs with attributes capturing object relationships. Feature engineering then produces a multi-dimensional input matrix of numerical features per activity. Testing showed that enriched features improved activity prediction accuracy, while an added LSTM layer enhanced the precision of remaining time prediction.

Galanti et al. \cite{galanti2023object} propose a framework integrating object interaction information through feature representation. Process views are built based on selected object types to reflect stakeholder-relevant KPIs. Object-Centric Event Logs are transformed into single-ID traces with attributes capturing object interactions via bridge events, enabling classical PPM techniques. Applying CatBoost to a dataset from an Italian utility provider improved prediction accuracy for five KPIs compared to traditional methods. The importance of object interactions was confirmed by metrics like MAE and F1 score, with Shapley values used to explain predictive mechanisms.

Adams et al. \cite{adams2023preserving} introduce a methodology for predicting next activity, timestamp, and remaining time while retaining graph structures in Object-Centric Event Logs. They compare Graph Neural Networks (GNNs), Graph Embeddings, and Flattened Event Logs. GNNs process graph-based inputs directly, embeddings transform graphs into vectors for regression models, and flattening applies traditional PPM techniques. Experiments show GNNs outperform other methods, retaining crucial graph-based information lost in flattening.

Galanti et al. \cite{galanti2023predictive} evaluate the trade-off between GNNs’ prediction performance and training time in PPM. Comparing Graph Convolutional Networks (GCNs), Gated Graph Neural Networks (GGNNs), LSTMs, and CatBoost across two processes and 21 KPIs, they find CatBoost significantly faster (8x) and more accurate with sufficient data. Unlike GNNs, which process graph structures natively, LSTMs and CatBoost rely on sequence encoding to handle object interactions efficiently.

\subsection{Classical Event Log}
This section provides an overview of PPM approaches based on classical event logs, categorising them according to the prediction task and method.

Di Francesco et al. \cite{di2016clustering} classify process traces as normal or deviant using a Labeling Function. Offline training involves encoding historical traces with frequency- and sequence-based methods, followed by clustering (model-based or DBSCAN) and supervised classification. Applied to hospital logs, the method balances accuracy and efficiency, reducing response times.

Tax et al. \cite{tax2017predictive} use LSTM networks for predicting next activity, timestamp, suffix, and remaining cycle time. Events are encoded with one-hot features and time-based attributes, processed by single-task and multi-task LSTM architectures. Evaluations on two datasets show LSTM outperforms specialized models but struggles with repetitive activity sequences.

Polato et al. \cite{polato2018time} tackle remaining time prediction in dynamic processes, comparing SVR, SVR with Transition Systems (TS), and Data-Aware Transition Systems (DATS), which combine classifiers and regressors. DATS performs well in stable scenarios, while SVR excels in dynamic contexts. However, suffix predictions using Damerau-Levenshtein similarity show random-like results.

Wang et al. \cite{wang2019outcome} propose an outcome-based prediction method using NLP techniques, transforming multi-class problems into binary tasks. Their Att-Bi-LSTM network combines bidirectional LSTMs and an attention mechanism to capture and prioritize contextual features. Evaluated on 12 datasets, it improves prediction accuracy with minimal efficiency trade-offs.

Park et al. \cite{park2019prediction} integrate Process Prediction Modeling (PPM) into resource scheduling in Business Process Management (BPM). Using an LSTM-based model, they optimize online resource allocation by predicting resource availability and activity times. Performance evaluation on real-life event logs shows improved Total Weighted Completion Time (TWCT), though increased Computation Time (CT) indicates computational challenges.

Camargo et al. \cite{camargo2019learning} propose a sequence prediction approach with pre-processing (embedding, scaling, n-grams), an LSTM-based model, and post-processing (random selection). Evaluations show strong performance for suffix prediction and robustness for long sequences but weaker next-event predictions. Remaining time predictions perform comparably, assessed with metrics like Damerau-Levenshtein distance and MAE.

Pasquadibisceglie et al. \cite{pasquadibisceglie2019using} introduce a CNN-based method for next activity prediction, transforming process traces into 2D images with channels for activity occurrences and elapsed time. Their approach outperforms LSTMs in next activity prediction but struggles with infrequent activities.

Mehdiyev et al. \cite{mehdiyev2020novel} approach next event prediction as a multi-classification task, using a sliding window for n-gram extraction and additional attributes. Their deep learning model, combining unsupervised pre-training and supervised fine-tuning, outperforms LSTM and other classifiers, excelling in rare event prediction via hyperparameter tuning and data balancing.

Park et al. \cite{park2020predicting} use traffic congestion prediction techniques for business process performance prediction. Their framework integrates annotated process models, spatial-temporal matrices, and an LRCN model combining CNNs and LSTMs. It outperforms baselines, achieving stable short- and long-term predictions by capturing spatial-temporal dependencies.

Pasquadibisceglie et al. \cite{pasquadibisceglie2020predictive} introduce PREMIERE, a 2D CNN-based approach for next activity prediction, transforming event logs into RGB images. Using Inception Blocks inspired by GoogLeNet, this method reduces overfitting and improves performance compared to LSTMs, RNNs, and other CNN-based models, demonstrating superior accuracy and precision.

Taymouri et al. \cite{taymouri2020predictive} propose a GAN-based framework for predicting next event labels and timestamps. Their method, using LSTMs in the GAN structure, outperforms three baselines in prediction accuracy and achieves competitive MAE for timestamps, showcasing the advantages of adversarial learning in process monitoring.

Bukhsh et al. \cite{bukhsh2021processtransformer} present ProcessTransformer, a self-attention-based architecture that captures long-range dependencies in event logs for next activity, event time, and remaining time prediction. The model achieves over 80\% weighted accuracy for next activity prediction, with lower MAE for event time and remaining time compared to baselines, validating the effectiveness of self-attention.

Kratsch et al. \cite{kratsch2021machine} compare ML and DL techniques for outcome-based prediction in process mining, showing that DL methods, particularly LSTM and DNN, outperform traditional ML methods, especially in handling complex event logs with high variance or imbalanced target variables.

Pasquadibisceglie et al. \cite{pasquadibisceglie2021multi} focus on next activity prediction using deep learning and multi-view learning. Their method, which incorporates multiple event perspectives, improves prediction accuracy but lacks interpretability.

Kotsias et al. \cite{kotsias2022predictive} apply process discovery algorithms with Q-learning for next activity prediction. Their approach, evaluated in the banking sector, requires around 500 training episodes to stabilize and achieve optimal performance.

Chiorrini et al. \cite{chiorrini2023multi} combine Instance Graphs (IG) and Deep Graph Convolutional Neural Networks (DGCNN) for next activity prediction. Their method uses the BIG algorithm to create Instance Graphs from event logs, incorporating a repair procedure to address anomalous traces. The enriched graphs are then processed by a DGCNN for prediction. Results show that BIG-DGCNN outperforms models like LSTM, CNN, MLP, and GCNN, particularly when enriched with additional process perspectives, leading to improved classification performance by capturing more complex process behaviors.

Oberdorf et al. \cite{oberdorf2023predictive} propose an "end-to-end enterprise process network" for predicting production disruption types by integrating event logs and departmental data. The framework involves five phases, including data acquisition, model design using a multi-headed neural network (MH-NN), and real-time prediction. Performance is evaluated using multi-class metrics such as accuracy and precision, with challenges like concept drift and longer prediction times compared to other methods.

Peeperkorn et al. \cite{peeperkorn2023can} test LSTM’s generalization in next event prediction by using leave-one-variant-out cross-validation. The model trains on prefixes and predicts the next event. Results show that while LSTM struggles with learning process structure, anti-overfitting measures help, though LSTM may create unseen variants.

Theis et al. \cite{9955412} extend Decay Replay Mining (DRM) for next event prediction by combining Petri Nets with neural networks. Their method includes three key extensions: Reachability Graph exploration, Masking Vectors, and feed-forward network adaptations. Results show improved prediction quality, especially for infrequent events, with applications in healthcare decision support.

Bousdekis et al. \cite{bousdekis2023modelling} apply Reinforcement Learning (RL) for next activity prediction. The process starts with process discovery on event logs, followed by computing transition probabilities. The Q-learning algorithm is used for training the RL agent, which predicts the next activity based on the process model. Results show Q-learning performs well for simple problems, while Deep Q Networks (DQN) are better for complex problems. Deep Q-Learning, which uses a neural network instead of a Q-table, requires more data but achieves better performance.

Razo et al. \cite{razo2023adjacency} introduce AXDP, a method for next event prediction that addresses overgeneralization in deep learning (DL). The approach uses adjacency matrices to preserve event sequence order. The method outperforms 75\% of baseline models, demonstrating the importance of maintaining event sequence integrity. However, it faces challenges with long event logs due to high computational time.

Rama et al. \cite{rama2023embedding} propose TACO, a Recurrent Graph Convolutional Process Predictor for next activity prediction. The approach combines spatio-temporal information from Petri nets and event sequences to predict the next activity. The model uses stacked GRNN and LSTM networks to handle both spatial and temporal dependencies. Evaluation on multiple datasets shows improved prediction accuracy by leveraging process structure in the model.

De Smith et al. \cite{de2023process} propose a method for predicting the global development of a process by representing event logs as multiple time series, focusing on Directly Follow Graphs (DFGs). The approach aggregates the event logs using equisize or equitemporal intervals and applies time series forecasting techniques (e.g., Holt Winter’s model, AR, ARIMA, GARCH, and VAR) to evaluate the process’s evolution. The method is tested on six real-life event logs, and the results are evaluated using mean absolute percentage error (MAPE).

Gunnarson et al. \cite{10045798} focus on suffix and remaining time prediction with LSTM, introducing the Complete Remaining Trace Prediction (CRTP) model. Unlike SEP-LSTM, which uses hallucinated predictions, CRTP-LSTM directly predicts the remaining trace and runtime by integrating all available information from previous events. The model outperforms others in predicting traces and runtime (using Levenshtein Similarity, MAE, and RMSE), with the inclusion of case and event features improving performance in most cases.

\section{Discussion}
The objective of this section is to provide a comprehensive and detailed examination of the previously analyzed approaches. The approaches will initially be grouped according to the type of event log, and subsequently, in order to account for the discrepancies between the various event log types, the following categories will be considered: Prediction Task and Method. 

\subsection{Event Log}
The proposals that were subjected to analysis were classified into two principal categories of event logs: The two main types of event log are object-centric and classical. The differentiation between these two categories resides in the presence of data concerning interactions between objects or case IDs, and the manner in which these are managed. 

In the case of object-centric event logs, two principal methodologies for the management of event logs have been identified. The initial approach entails the flattening of the logs to align them with the characteristics of a Classical Event Log, followed by the incorporation of interaction data. Proposals such as those presented in \cite{gherissi2022object,galanti2023object} adopt this approach, applying state-of-the-art methods for classical event logs. The second category preserves the graph structure of object-centric event logs by leveraging control-flow data, employing predictive models such as graph neural networks (GNN) and graph embedding. As demonstrated by studies such as \cite{adams2023preserving} and \cite{galanti2023predictive}, the utilisation of graphs can enhance prediction accuracy. Nevertheless, the deployment of straightforward models with meticulous feature engineering can also yield favourable outcomes with reduced training time.

As indicated in \cite{DiFrancescomarino2022}, the selection of process perspective for analysis has a profound effect on how data are employed. The incorporation of supplementary data from multiple perspectives, including contextual information, has the potential to enhance the accuracy of predictions. While some approaches concentrate exclusively on control-flow behaviour, others supplement the event log with data from a variety of perspectives. The integration of supplementary data has been demonstrated to result in enhanced predictive outcomes \cite{polato2018time,pasquadibisceglie2019using,pasquadibisceglie2020predictive,kratsch2021machine,pasquadibisceglie2021multi,chiorrini2023multi}.

\subsubsection{Prediction Task}
The Prediction Task is related to the process variable that is being predicted. This may be the outcome, the numeric value, or the next event, as was discussed previously. A prediction may be made with regard to the conclusion of a case or a particular future point. The selection of the prediction algorithm and evaluation metrics is frequently contingent upon the specific task at hand. 

In Object-Centric Event Log approaches, prediction tasks are generally focused on next event or numeric value prediction \cite{adams2023preserving,gherissi2022object}, with some work on KPI prediction \cite{galanti2023predictive,galanti2023object}. The prediction task is less central to the approach structure in these studies.

In Classical Event Log approaches, a distinction is made between predicting the next event (with an activity label and timestamp) and predicting the case until completion (with suffix and remaining time). Most proposals focus on next event prediction, with some extending their work to suffix or remaining time prediction \cite{tax2017predictive,camargo2019learning,bukhsh2021processtransformer}. Few studies, such as \cite{de2023process,10045798}, focus on suffix prediction. De Smeth et al. \cite{de2023process} approach process prediction using time series techniques, avoiding next event prediction. Gunnarson et al. \cite{10045798} tackle suffix prediction by addressing the hallucination effect, using only observed log data for more accurate predictions.

For outcome prediction, some studies focus on binary outcome prediction \cite{di2016clustering}, while others transform it into multiple binary prediction problems \cite{wang2019outcome}, and others address it in different ways \cite{park2020predicting,kratsch2021machine,oberdorf2023predictive}.

\subsubsection{Method}
As discussed, prediction methods in PPM fall into model-based (process-aware) and non-process-aware approaches, both relying on event log encoding for prediction. This section reviews common models, emphasizing performance, explainability, interpretability, and training time. It distinguishes between Object-Centric and Classical Event Logs, and then categorizing methods into ML-based and DL-based solutions.

The most commonly used method in Object-Centric Event Logs is Graph Neural Networks (GNNs), which have shown improved performance. However, recent studies \cite{galanti2023predictive} highlight that GNNs often require longer training times relative to the performance gains they offer. By focusing on feature engineering, Object-Centric Event Logs' properties can be incorporated without relying on GNNs. The second most popular method is Long Short-Term Memory (LSTM), as used in \cite{gherissi2022object} and \cite{galanti2023object}. These studies focus on next event prediction and remaining time using both filtered and enriched event logs.

In Classical Event Logs, LSTMs are widely used, recognized for handling long dependencies. They are applied across several prediction tasks, including next activity prediction \cite{tax2017predictive,camargo2019learning,taymouri2020predictive,pasquadibisceglie2021multi,peeperkorn2023can}, outcome prediction \cite{wang2019outcome,park2019prediction,kratsch2021machine}, and remaining time prediction \cite{10045798}. LSTM architectures are often modified or combined with other techniques to improve performance \cite{wang2019outcome,taymouri2020predictive,rama2023embedding,10045798}. Other methods like CNNs \cite{pasquadibisceglie2019using,pasquadibisceglie2020predictive} and GNNs \cite{chiorrini2023multi} are also used, with GNNs leveraging multi-perspective enriched instance graphs.

Transformers \cite{bukhsh2021processtransformer} offer improved predictions but require longer training times and provide less explainable results.

Machine learning models like Decision Trees (DT), Random Forests (RF), Support Vector Machines (SVM), and Q-Learning \cite{kotsias2022predictive,bousdekis2023modelling} are also utilized. These models provide good performance and more explainable results compared to DL-based methods.

While methods like Transformers and Deep Learning (DL) approaches generally outperform ML models in terms of prediction accuracy, they show challenges. Training time is a significant issue, especially with networks like GNNs and Multi-Headed DNNs, which require extensive training compared to other methods. Additionally, LSTMs can suffer from over-generalization, producing behaviors not observed in the event log \cite{peeperkorn2023can,razo2023adjacency}. Finally, explainability is a concern with DL-based methods and Transformers, as these models often lack the capability to provide clear explanations for their predictions.

\section{Conclusion}
The objective of this paper is to present a systematic literature review (SLR) of the techniques employed in the domain of Predictive Process Monitoring (PPM). The increasing adoption of sophisticated information systems, with the advent of big data analytics, has led to a heightened importance of predictive process monitoring (PPM) in organisational contexts, anticipating future process behaviours. These techniques employ historical event logs recorded by information systems to forecast future process outcomes, thereby enabling organisations to make informed decisions in response to evolving conditions.

After introducing key concepts on Event Logs and PPM techniques, we outline the main contribution of our work in relation to existing reviews. To this end, we provide a related work section that examines the objectives and methodologies of current reviews in the field.

Then we present a structured and detailed research methodology to analyze and categorize the proposed PPM approaches found in the literature. We classify them based on the type of Event Log used, distinguishing between Object-Centric Event Logs and Classic Event Logs. This distinction is the key contribution of our work, as it is the first review to explicitly differentiate between these Event Log types. This review emphasizes the limitations of current research on Object-Centric Event Logs compared to the advancements achieved with Classic Event Logs. Through a thorough discussion, we classify the examined approaches by their characteristics and highlight notable features. 

From our analysis, we found that approaches using Object-Centric Event Logs primarily focus on managing interaction information between objects, facing challenges in balancing prediction accuracy with computational efficiency. In contrast, Classic Event Log approaches focus more on enhancing prediction accuracy through techniques like advanced encoding methods and robust prediction algorithms, such as neural networks and transformers.

The findings presented here offer a starting point for a more focused research approach, supporting the improvement of the structure and methodology within the research field, creating a valuable resource for new researchers. 

As part of our future work, we plan to further investigate the impact of event log characteristics such as data encoding, perspective, availability, variability, feature engineering and complexity on prediction accuracy, and then identify a subset of the analysed methods based on the identified dimensions and perform a benchmark analysis of the selected techniques.

\bibliographystyle{splncs04} \bibliography{mybibliography}

\end{document}